\documentclass[sigconf,screen]{acmart}

\AtBeginDocument{%
  }

\setcopyright{acmlicensed}
\copyrightyear{2026}
\acmYear{2026}
\acmDOI{XXXXXXX.XXXXXXX}
\acmConference[FSE 2026]{ACM International Conference on the Foundations of Software Engineering}{July 05--09,
  2026}{Montreal, Canada}
\acmISBN{978-1-4503-XXXX-X/2018/06}

\begin{document}

\title{Toward Scalable Automated Repository-Level Datasets for Software Vulnerability Detection}

\author{Amine Lbath}
\author{Supervisor: Prof. Massih-Reza Amini}
\email{first.last[at]nist.gov;first.last[at]univ-grenoble-alpes.fr}
\affiliation{%
  \institution{Univeristy of Grenoble}
  \city{Grenoble}
  \country{France}
}

\renewcommand{\shortauthors}{A. Lbath}

\begin{abstract}
Software vulnerabilities continue to grow in volume and remain difficult to detect in practice.
Although learning-based vulnerability detection has progressed, existing benchmarks are largely function-centric and fail to capture realistic, executable, interprocedural settings.
Recent repo-level security benchmarks demonstrate the importance of realistic environments, but their manual curation limits scale.
This doctoral research proposes an automated benchmark generator that injects realistic vulnerabilities into real-world repositories and synthesizes reproducible proof-of-vulnerability (PoV) exploits, enabling precisely labeled datasets for training and evaluating repo-level vulnerability detection agents.
We further investigate an adversarial co-evolution loop between injection and detection agents to improve robustness under realistic constraints.
\end{abstract}




\maketitle

\section{Problem Statement and Motivation}

The number of publicly disclosed software vulnerabilities has increased steadily over the past decade, with recent years reporting tens of thousands of new CVEs annually~\cite{nistNVD}.
At the same time, modern software systems have grown in size and complexity, increasingly relying on large, multi-module codebases and rapid development cycles.
This trend is further amplified by the widespread adoption of AI-assisted coding tools, which have been shown to generate insecure code patterns and to exacerbate existing security risks when used without rigorous analysis~\cite{veracode2025genai}.
As a result, effective vulnerability detection must operate at the \emph{codebase level}, where vulnerabilities often emerge from interactions across functions, files, and configuration boundaries rather than from isolated functions.

\paragraph{Why current benchmarks are the bottleneck.}
A large fraction of ML/LLM vulnerability detection research still reduces the task to \emph{function-level binary classification}.
Recent analyses show this formulation is often ill-defined: many functions cannot be labeled vulnerable/benign without their calling context, and high scores may be driven by spurious correlations rather than true reasoning about security~\cite{risse2025topscore}.
Likewise, evaluations on traditional datasets can dramatically overestimate real-world effectiveness; more realistic settings reveal large performance drops~\cite{ding2024primevul,chakraborty2024realvul}.

\paragraph{Repo-level datasets exist, but realism and scalability remain limited.}
Several recent efforts move beyond single functions by mining repository context from CVE/fix histories.
ReposVul and VulEval provide interprocedural context and large-scale mining, but depend on proxy labels and evaluation setups that are not always executable end-to-end~\cite{wang2024reposvul,wen2024vuleval}.
AI agent-focused benchmarks, such as BountyBench and CVE-Bench, explicitly target realistic environments and security-agent tasks (detect/exploit/patch) with reproducible setups, but require substantial manual effort and remain limited in scale, which constrains their use for training~\cite{zhang2025bountybench,zhu2025cvebench}.
Meanwhile, specialized benchmarks highlight that repo-level security tasks are difficult even for strong AI agents, motivating the need for larger, training-ready datasets~\cite{zhang2025bountybench, guo2025frontier}.

\paragraph{Gap.}
Existing benchmarks lack a scalable mechanism to generate \emph{repository-level}, \emph{buildable}, and \emph{executable} vulnerability instances with precise labels and reproducible proof-of-vulnerability artifacts.
Recent work on automated vulnerability injection at the function level, including our own (AVIATOR), demonstrates that controlled injection can produce realistic and high-quality security data~\cite{lbath2025aviator}.
This research builds on that insight by extending vulnerability injection to the repository scale and systematically pairing injected vulnerabilities with executable proofs, enabling realistic training and benchmarking of repo-level vulnerability detection agents.

\section{Research Questions}
Recent LLM-based and agentic approaches demonstrate emerging capabilities for software vulnerability detection, yet their performance remains limited in realistic, repository-level settings.
Several recent studies explicitly identify the lack of large-scale, executable, and accurately labeled codebase-level vulnerability datasets as a key obstacle to building stronger systems and to benchmarking them reliably~\cite{guo2025frontier,zhang2025bountybench}.
To address this gap, this doctoral research is guided by the following questions:
\begin{itemize}
  \item \textbf{RQ1 (Benchmark quality).} Can AI agentic workflows inject \emph{realistic} vulnerabilities into real repositories while preserving build/test validity and yielding reproducible PoVs?
  \item \textbf{RQ2 (Training utility).} Does training on a large, codebase-level, dataset improve generalization and robustness of vulnerability detection models compared to training at function-level?
  \item \textbf{RQ3 (Adversarial co-evolution).} 
    Can an adversarial co-evolution between vulnerability injection and detection agents improve robustness in repository-level vulnerability detection?
\end{itemize}

\section{Proposed Approach}
The core contribution is an \textbf{automated benchmark generator} driven by a workflow of AI agents.
The design is \emph{human-expert inspired}: it mirrors how security engineers (i) set up a target, (ii) form a vulnerability hypothesis, (iii) implement a minimal-but-plausible change, and (iv) validate exploitability with a PoV.

\paragraph{\textbf{Phase A: Target selection and executable harness}}
Given a repository, the system automatically:
(i) builds the project in a containerized environment, (ii) discovers or synthesizes a test harness (existing tests, fuzz harnesses, or minimal driver programs), and
(iii) establishes invariants (tests passing, sanitizer baseline, API contracts) used to reject problematic injections.
This step aims to ensure every benchmark item is \emph{runnable} and evaluation is \emph{reproducible}.

\paragraph{\textbf{Phase B: Vulnerability Injection with Multi-Agent Control}}
To identify realistic injection points at repository scale, we leverage \emph{structural and semantic signals} extracted from the codebase.
In particular, a \textbf{CodeQL-guided analysis} \cite{codeql_vustinov2016} is used to mine candidate locations based on dataflow patterns, and security-relevant sinks and sources, providing a principled approximation of how human experts reason about vulnerability placement.

A set of cooperating agents then performs the injection:
\begin{itemize}
  \item \textbf{Planner agent} selects a vulnerability class (e.g., CWE family) and candidate injection sites suggested by CodeQL queries (e.g., incomplete validation paths, risky sink reachability).
  \item \textbf{Implementer agent} introduces minimal, cross-file changes that activate the vulnerability only along specific execution paths, favoring interprocedural flaws.
  \item \textbf{Reviewer agent} enforces realism constraints (coding style, plausible developer intent, avoidance of synthetic signatures).
  \item \textbf{Verifier agent} rebuilds the project and executes tests to ensure that only intended behaviors are affected.
\end{itemize}

This combination of static query guidance and agentic editing enables scalable, human-inspired vulnerability injection while preserving repository integrity.

\paragraph{\textbf{Phase C: Proof-of-vulnerability synthesis and trace artifacts}}
For each injected vulnerability, an exploit-generation agent attempts to produce a \textbf{PoV}:
\begin{itemize}
  \item For memory safety: a triggering input with sanitizer/crash signature and a minimal reproducer.
  \item For logic/web-style flaws: an end-to-end request sequence and observable security violation.
\end{itemize}
The system records \emph{executable traces} (stack traces, crashing inputs, dependency slices, and minimal call paths) to support evaluation of localization and explanation quality.
Each benchmark item includes: original code, vulnerable commit, container recipe, harness, PoV, and structured labels (CWE, affected components, trigger path).

\paragraph{\textbf{Phase D: Training and benchmarking repo-level detection agents}}
We leverage the generated corpus to train \textbf{vulnerability detection agents} that operate at repository scope:
tool-augmented retrieval over the codebase, iterative localization, hypothesis refinement, and validation (e.g., reproducing the PoV or generating a patch). By providing executable ground truth and cross-file context, the benchmark enables evaluation of detection systems in settings that more closely reflect real-world software engineering workflows.

\paragraph{\textbf{Phase E: Adversarial co-evolution loop}}
Finally, we propose a novel \textbf{adversarial learning framework} consisting of two competing AI models: a Vulnerability Injector and a Vulnerability Detector. Through competitive co-evolution, these models will iteratively improve each other's capabilities, with the injector learning to generate realistic code vulnerabilities while the detector becomes increasingly skilled at identifying them.

\section{\textbf{Evaluation Plan and Expected Contributions}}
\textbf{Benchmark validation} will measure: (i) build/test pass rate after injection, (ii) PoV reproducibility, (iii) realism via expert audit on a stratified sample, and (iv) diversity (CWE distribution, multi-file depth).
\textbf{Training utility} will compare detectors trained with/without injected repo-level data, tested on external repo-level benchmarks where possible (e.g., BountyBench/CVE-Bench-style tasks)~\cite{zhang2025bountybench, zhu2025cvebench}.
\textbf{Contributions} include: (1) a novel scalable repo-level vulnerability benchmark generator with PoVs, (2) a dataset to train repo-level vulnerability detection agents, and (3) an adversarial co-evolution methodology to continuously improve both injection and more importantly  detection.

\bibliographystyle{ACM-Reference-Format}
\bibliography{document}

\end{document}